\documentclass[
	11pt,	
	a4paper,	
	twocolumn
]{article}

\usepackage[a4paper,left=2.5cm,right=2.5cm,top=2.5cm,bottom=2.5cm]{geometry}

\usepackage[utf8]{inputenc}
\usepackage[english]{babel}
\usepackage[T1]{fontenc}
\usepackage{amsmath}
\usepackage{amsfonts}
\usepackage{amssymb}

\usepackage{graphicx}
\usepackage{float}
\usepackage{color}
\usepackage{hyperref}
\usepackage{subcaption}
\usepackage{trfsigns}
\usepackage{import}
\usepackage{multirow}
\usepackage{authblk}

\title{The Combination of Several Decorrelation Methods to Improve Acoustic Feedback Cancellation}
\author[1]{Klaus Linhard}
\author[2,*]{Philipp Bulling}

\affil[1]{Christian-Albrechts-Universität zu Kiel, Germany}
\affil[2]{Hochschule Esslingen, Germany}

\affil[*]{Corresponding author: \href{mailto:philipp.bulling@hs-esslingen.de}{philipp.bulling@hs-esslingen.de}}

\newcommand{\sub}[1]{_{\mathrm{#1}}} 						
\newcommand{\abs}[1]{\left| {#1} \right|}					
\newcommand{\norm}[1]{\lVert {#1} \rVert}					

\newcommand{\vect}[1]{\boldsymbol{#1}}						

\newcommand{\h}{\vect{h}}									
\newcommand{\hEst}{\hat{\vect{h}}}							

\begin{document}
\maketitle

\section*{Abstract}
This paper extends an acoustic feedback cancellation system by incorporating multiple decorrelation methods. The baseline system is based on a frequency-domain Kalman filter implemented in a multi-delay structure. The proposed extensions include a variable time delay line, prediction, distortion compensation, and a simplified reverberation model. Each extension is analyzed, and a practical parameter range is defined.

While existing literature often focuses on a single extension, such as prediction, to describe an optimal system, this work demonstrates that each individual extension contributes to performance improvements. Furthermore, the combination of all proposed extensions results in a superior system. The evaluation is conducted using publicly available datasets, with performance assessed through system distance metrics and the objective speech quality measure PSEQ.

\section{Introduction}
This work addresses the challenge of acoustic feedback cancellation in speech communication systems, with a focus on applications such as in-car passenger communication \cite{14_Schmidt2006}. The primary objective is to introduce and validate useful extensions to a baseline approach, demonstrating their effectiveness in improving system performance.

The foundational feedback cancellation system employs a frequency-domain Kalman filter. To enhance stability and adaptation, several decorrelation methods are proposed and analyzed. A key contribution of this work is the successful combination of these methods, yielding a robust and improved system.

Validation is conducted through two complementary approaches: objective speech quality assessments using the Perceptual Evaluation of Speech Quality (PESQ) metric \cite{17_Hu2008}, and technical evaluations based on system distance metrics and overflow analysis. The remainder of this introduction focuses on the technical and mathematical challenges inherent in feedback cancellation. Specifically, it provides a concise overview of the frequency-domain Kalman filter structure and examines two critical adaptation issues: bias and convergence speed.

\subsection{Structure of an Acoustic Feedback System}
\label{sec:structure}
\begin{figure}
  \centering
  \includegraphics{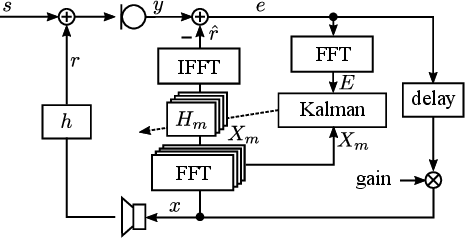}
  \caption{Structure of the closed loop feedback system showing the integration of a Kalman filter into a multi-delay frequency domain filter.}
  \label{fig:block_diagram}
\end{figure}

Fig.\,\ref{fig:block_diagram} depicts a simplified structure of an acoustic feedback system. A microphone captures the signal $y = s + r$, which comprises the original speech signal $s$ and the room echo $r$. The room's impulse response is denoted by $h$. The loudspeaker signal $x$ is the origin of the room echo $r$. From $y$, the room echo estimate $\hat{r}$ is subtracted, yielding the error signal $e = y - \hat{r}$. Before reaching the loudspeaker, $e$ undergoes a delay and amplification with a loop gain of $g$. The room echo $r$ is estimated in the frequency domain after applying Fast Fourier Transforms (FFTs), multiple adaptive filters $H_m$, and an inverse FFT (IFFT). We employ the so-called multi-delay frequency domain least mean squares error structure (MD-FLMS), where the complete filter is portioned into several parts $m = 1, 2, \ldots, M$, as originally proposed in \cite{5_Soo1990}. A frequency domain Kalman filter is integrated into this structure, as demonstrated in \cite{6_Enzner2006, 7_Kuech2014, 11_Bernardi2017}.

\noindent
For acoustic feedback cancellation in rooms, such as a car cabin, the impulse response $h(k)$ can have a length of 1000 samples or more, assuming a sampling frequency of 16\,kHz. To address this, we partition the signal $x$ into segments of length $N=512$, with half overlap between consecutive segments. Consequently, the partitions of the useful impulse response have a length of 256. Covering the 1024 samples of the room impulse response, requires $M=4$ partitions. These partitioning parameters serve as the basis for the remainder of this paper.

It is well established that the Kalman filter is more robust in tracking a variable room impulse response, as it incorporates provisions for tracking \cite{6_Enzner2006}. However, we will not delve into the tracking capabilities here, instead focusing on two fundamental problems:

\begin{itemize}
\item bias reduction and
\item convergence speed improvement.
\end{itemize}

Our approach involves reducing the autocorrelation of $x$ and the cross-correlation between $x$ and $s$, which are essential steps in addressing these problems.

\subsection{The Problem of Reducing Bias}
The error criterion used to derive the optimum estimate $\hat{h}$ of the room impulse response $h$ is the minimum mean square error, expressed in statistical notation as the expectation value $E\left\{e^2\right\}\rightarrow \min$. Instead of directly deriving a Kalman filter model, e.g., \cite{6_Enzner2006}, we consider a simpler derivation based on the aposteriori least mean error criterion
\begin{equation}
\varepsilon(k)= s(k)+ \left(\h(k)-\hEst(k)\right)^T \vect{x}(k)                                                         
\end{equation}
where $\hEst(k)$ is the updated filter coefficients vector, as described in \cite{8_Puder2017}. Minimizing $E\left\{e^2\right\}$ with respect to $\hEst(k)$ yields
\begin{equation}
\hEst\sub{opt}=\h+\h\sub{bias}= \h+ \vect{R}_{xx}^{-1}  \vect{r}_{xs}.
\end{equation} 
The estimated optimum impulse response $\hEst\sub{opt}$ is a sum of two impulse responses: the true impulse response $\h$ and the impulse response of the bias, $\h\sub{bias}$. Here $\vect{R}_{xx}$ is the autocorrelation matrix of $\vect{x}$ and $\vect{r}_{xs}$ is the cross-correlation vector between $\vect{x}$ and $s$. The second part of the impulse response, acts as a predictor, subtracting the predictable components of $s(k)$. This whitening of the speech signal $s$ results in the cancellation of colored noise and especially periodic voiced speech parts, which is undesirable.

To reduce the bias, we will first examine the fixed time delay (a linear processing step) and then consider several non-linear processing steps:

\begin{itemize}
\item non-linear distortion in the time domain
\item time-varying delay line
\item reverberation model
\end{itemize}

The known methods of frequency shifting and phase modulation are discussed in \cite{9_Guo2012} and \cite{16_Linhard2025}. Within the class of phase/frequency effects, we only employed the time-varying delay line, also known as the vibrato audio effect. The advantage of this approach lies in its simple and straightforward implementation. We did not include the method of noise injection (e.g., \cite{4_Valin2016}) in our selection, as it typically requires a significantly higher implementation effort compared to the suggested simple methods.

\subsection{The Problem of Low Convergence Speed}
It is well established that time-domain LMS algorithms suffer from the eigenvalue spread of the autocorrelation matrix of the input signal $x$, resulting in slow convergence speeds for colored input signals. To address this issue, prewhitening of the input signal using predictors has been suggested to increase convergence speed, as discussed in \cite{10_Kuehl2017,11_Bernardi2017}. Frequency-domain algorithms, such as FMLS and MD-FLMS, as well as frequency-domain Kalman filters, provide a frequency-selective power normalization that leads to signal whitening and, consequently, significant speed improvements. Even with frequency-domain approaches, the addition of predictors remains useful, as demonstrated in \cite{10_Kuehl2017,11_Bernardi2017}. For positioning the predictors, we primarily follow the approach outlined in \cite{10_Kuehl2017}, but we need to adapt and extend this approach to accommodate the multi-delay structure.

\section{Performance Evaluations}
\label{sec:performance}
For each method, we will first present results for one phonetically balanced male speech sentence. In later sections, we will provide additional results using a speech database. The room impulse response used throughout this paper is an example of length 1024. The parameters of the MD structure have already been specified in Sec.\,\ref{sec:structure}. For the Kalman filter, we choose the following parameters: $\alpha=1$; $\gamma=0.1$; $\delta=0.2$; $A=0.99999$, as described in \cite{7_Kuech2014}. We apply loop gain values of $g = 0$, 6, 12, or 30\,dB, as shown in Fig.\,\ref{fig:block_diagram}. With $g = 0\mathrm{\,dB}$, we adjusted the acoustic coupling between the loudspeaker and microphone to approximately -10\,dB, meaning that the signal level from the loudspeaker signal (feedback) is about -10\,dB lower than the level of the original speech from the microphone. Although it is essential to consider environmental noise in practical implementations, which can disturb adaptation, this issue is not addressed in this paper.

The system distance, denoted as $\mathit{sd}(l)$, depends on the time block index $l$, the room impulse response vector $\h$, and the actual impulse response estimate $\hEst_l$. It is calculated as
\begin{equation}
\mathit{sd}(l)= \frac{\norm{\h-\hEst_l}}{\norm{h}}, 
\label{eq:sysDist}
\end{equation}
where $\norm{\cdot}$ represents the L2-norm. Due to the half overlap, each block ($N=512$) contains 256 new data samples. To assess the speed of adaptation, we propose using an early system distance, $\mathit{sd}_5$, and a late system distance (after convergence), $\mathit{sd}_{20+}$. $\mathit{sd}_5$ is measured as the average of $\mathit{sd}(l)$ from Eq.\,\ref{eq:sysDist} near 5\,sec, i.e., over the time interval $[4\ 6]$\,sec, and $\mathit{sd}_{20+}$ is an average after 20\,sec (for several seconds, or until the end of the speech).

We selected the PESQ method, as described in \cite{17_Hu2008}, as an objective speech evaluation metric. PESQ models the mean opinion scores (MOS) that cover a scale from 1 (bad/very annoying) to 5 (excellent/imperceptible), with intermediate scores of 2 (poor/annoying), 3 (fair/slightly annoying), and 4 (good/perceptible but not annoying).

The results obtained using PESQ depend on the loop gain $g$, which is set to 0, 6, 12, or 30\,dB. Short, locally unstable phases may occur, for example, during the adaptation-start phase, particularly in the case of a high loop gain of 30\,dB. To mitigate this, we use a ramp to linearly increase $g$ from 0\,dB to the final value over a period of 0, 1, 2, or 10\,sec, depending on the final gain value.

The evaluation is divided into two parts. The first part, covering Sec.\,\ref{sec:bias_reduction} and Sec.\,\ref{sec:convergence_speed}, focuses on finding a useful range of parameters based on specific quality measures and system distance convergence curves, using only one speech example.

The second part, presented in Sec.\,\ref{sec:combination_decorrelation}, uses a speech database for evaluation. To ensure a secure procedure and avoid system breakdown due to instability, we implement a hard limiter to clip signal amplitudes at +6\,dB and count the relative amount of clippings, referred to as overflows (in [\%]). We will provide these overflow counts only for the challenging 30\,dB gain case, as the overflow count is generally small (usually close to 0) for smaller gains. 

\section{Methods of Bias Reduction}
\label{sec:bias_reduction}
We have previously introduced the bias reduction methods, and now we will delve into the specifics of each individual method.

\subsection{Bias Reduction with Fixed Time Delay}
In our approach, a fixed time delay is a direct consequence of the multi-delay structure. The delay is equivalent to the size of one partition of the impulse response, which, with our parameter setting, is 256 samples. A time delay of 256 samples already provides a significant amount of signal decorrelation. As discussed in \cite{16_Linhard2025}, this effect can be quantified. We consider this delay as a structure-inherent advantage throughout the paper, and it will be accepted as a baseline for the subsequent bias reduction methods.

\subsection{Bias Reduction with Variable Time Delay (Vibrato)}
In \cite{9_Guo2012}, the combination of frequency shift and phase modulation was proposed to decorrelate signals $s$ and $x$. The implementation was based on a complex subband approach. Later, in \cite{16_Linhard2025}, a DFT filterbank was suggested as a flexible tool to realize frequency shift, phase modulation, or variable time delay lines. Essentially, all these methods can be viewed as phase modulation. The variable time delay line is known as a natural audio effect, the vibrato, or in an extended version, the chorus \cite{12_Zoelzer2011}. In this paper, we will focus on the variable time delay only. We will realize the vibrato effect in the time domain. Signal $x$ is the input of a tapped delay line, and the output is one of the taps with the time index of the tap modulated from a sine signal. The necessary output values between two taps are interpolated using a 1st-order allpass filter. An appropriate algorithm is shown, for example, in \cite{12_Zoelzer2011}. Several other more sophisticated algorithms are known, such as \cite{13_Smith2022}. Important parameters of the variable delay line are the maximum delay and the frequency of the modulation. From \cite{16_Linhard2025}, we know that a maximum delay of about 1 to 2\,msec and a modulation frequency of about 1 to 2\,Hz deliver good results.
\begin{figure}[t]
\centering
\includegraphics[width=\linewidth]{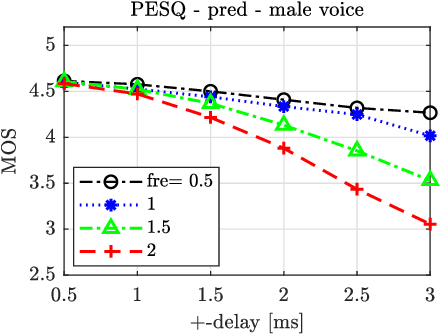}
  \caption{MOS values for a variable time delay line (vibrato effect) depending on modulation frequency and max. delay time.}
  \label{fig:vibrato_mos}
\end{figure}\noindent 
Fig.\,\ref{fig:vibrato_mos} presents the MOS (Mean Opinion Score) quality measure of the proposed time domain vibrato realization, measured using a phonetically balanced male speech sentence (without the feedback system).
\begin{figure}[t]
\centering
\includegraphics[width=\linewidth]{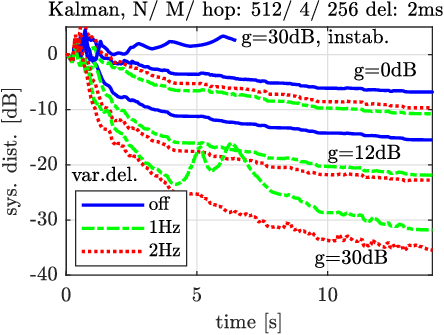}
  \caption{Convergence of the system distance for $g=0$, 12 and 30\,dB. Max. delay is $\pm 2$\,msec and modulation frequency is  either off,  1, or 2\,Hz.}
  \label{fig:vibrato_sd}
\end{figure}\noindent
Fig.\,\ref{fig:vibrato_sd} depicts the system distance of the convergence phase for gain values of $g=0$, 12, and 30\,dB. We fixed the maximum delay to $\pm 2$\,msec. In the 30\,dB gain case, without any extension, the system is unstable, resulting in a system breakdown after approximately 6\,sec. With 1\,Hz modulation and $g=30$\,dB, there is some uncertainty around 5 to 6\,sec, but the system still exhibits continuing convergence. With 2\,Hz modulation, we observe a reduced MOS value (see Fig.\,\ref{fig:vibrato_mos}) but an improved convergence curve. The variable delay line significantly enhances the results.

In contrast to the realization as a phase synthesizer with a DFT filterbank in \cite{16_Linhard2025}, we implemented the variable time delay line (vibrato) in the time domain. Our system structure (see Fig.\,\ref{fig:block_diagram}) already contains a block named "delay," which can be extended with a "variable time delay." Input and output signals of this delay are time domain signals. Unlike the frequency domain solution \cite{16_Linhard2025}, we can work with an implementation latency of only about 2\,msec (to implement a flexible delay of max 2\,msec) and do not need to consider extra FFT segmentation delays. However, a disadvantage is that we do not obtain all the flexible phase modifications of a phase synthesizer \cite{16_Linhard2025}.

\subsection{Bias Reduction with Non-linear Distortion}
Already in the 90ies, it was suggested to use the half-wave rectifier to improve the acoustic echo cancellation of stereo signals \cite{1_Benesty1997}. The basic idea was to include a non-linear function into the stereo channels to reduce the interchannel coherence and thus support the estimation of a unique solution of the two room impulse responses. The non-linear distortion should not be noticeable due to self-masking of the speech signal. It was further shown in \cite{Bulling2016} that non-linearities can even improve speech intelligibility. We will compare four non-linear functions:
\linebreak
1. Half-wave rectification:
\begin{equation}
y_1(k) = 0.5 \cdot (x(k) + \abs{x(k)}),
\label{eq:halfwave}
\end{equation}
2. Signed square:
\begin{equation}
y_2(k) = x(k) \cdot \abs{x(k)},
\label{eq:signedsquare}
\end{equation}
3. Half-wave rectification and signed square:
\begin{equation}
y_3(k) = 0.5 \cdot (y_1(k) + y_2(k)),
\label{eq:halfwavesignedsquare}
\end{equation}
4. Smoothed half-wave rectification:
\begin{equation}
y_4(k) = 0.5 \cdot (x(k) + \sqrt{x(k)^2 + c^2}),
\label{eq:smoothedhalfwave}
\end{equation}
with
\begin{eqnarray}
c &=& 0.65 \sqrt{\sigma_x^2(k)} \label{eq:smoohhalf1}\\
\sigma_x^2(k) &=& (1 - \beta) \sigma_x^2(k-1) + \beta x^2(k). \label{eq:smoohalf2}
\end{eqnarray}
Note that the smoothed half-wave rectification includes a smoothing factor $c$ that is adapted based on the signal variance $\sigma_x^2(k)$.

Eq.\,\ref{eq:halfwave} generates even harmonics only, while the signed square Eq.\,\ref{eq:signedsquare} produces odd harmonics only. We propose combining odd and even harmonics as curve $y_3$, Eq.\,\ref{eq:halfwavesignedsquare}. This curve is a promising candidate due to its dense grid of even and odd harmonics.

The smoothed half-wave rectifier Eq.\,\ref{eq:smoothedhalfwave} is a modified version of the standard half-wave rectifier from Eq.\,\ref{eq:halfwave}, featuring a soft knee \cite{2_Morgan2001,4_Valin2016,10_Kuehl2017}. By avoiding hard cuts at the amplitude zero crossings, this approach should result in lower distortion. The radius of the knee is controlled by the additive offset $c$, which is made adaptive by depending on the signal's standard deviation $\sigma_x$. Parameter $\beta$ in Eq.\,\ref{eq:smoohalf2} is used to estimate the variance (we use $\beta=0.005$ with a sampling frequency of 16\,kHz). The factor 0.65 is used in literature \cite{2_Morgan2001,4_Valin2016} without derivation, and we reuse it here. A signal with high variance receives a larger radius compared to a signal with lower variance.

\begin{figure}
\centering
\begin{subfigure}{\linewidth}
  \includegraphics[width=\linewidth]{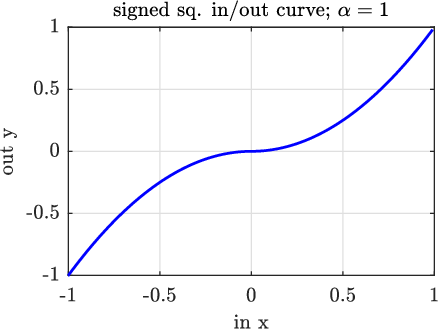}
\end{subfigure}
\begin{subfigure}{\linewidth}
  \includegraphics[width=\linewidth]{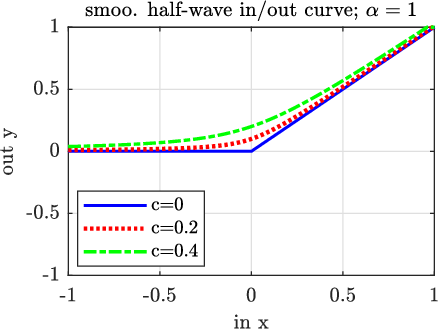}
\end{subfigure} 
  \caption{Examples for non-linear functions. Top: Signed square; Bottom: Smoothed half-wave rectification with different but fixed values of $c$.}
  \label{fig:nonlinear_functions}
\end{figure}

Fig.\,\ref{fig:nonlinear_functions} displays two examples of non-linear functions, with the input range of signal $x$ limited to the interval $[-1\ 1]$.
\noindent 

\begin{figure}
\centering
\includegraphics[width=\linewidth]{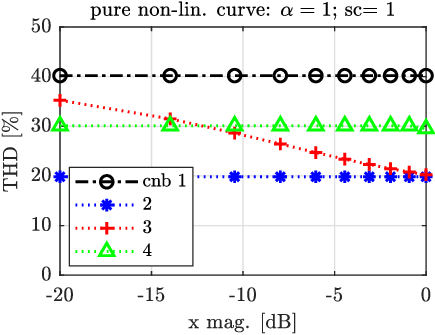}
  \caption{THD comparison of curves 1 to 4, for input magnitude range of $x$ from  -20\,dB to 0\,dB (0\,dB represents magnitude 1;  $\alpha=1$, $\mathit{sc}=1$, cf. Eq.\,\ref{eq:scaledmix}).}
  \label{fig:thd}
\end{figure}

A well-established measure of non-linearity is the total harmonic distortion (THD). The input $x$ for a THD measure is a sinusoid of a certain frequency, the first harmonic. The THD value is the square root of the ratio of the power sum of all higher harmonics to the power of the first harmonic, usually given in \%. For our small collection of non-linear functions, we observe the remarkable feature that in Fig.\,\ref{fig:thd}, curves no. 1, 2, and 4 show a constant THD independent of the magnitude of input $x$, while $y_3$ provides higher THD for smaller input levels. In our comparison, $y_1$ exhibits the highest distortion.
\noindent

To adjust a flexible effect, it is a common procedure in audio effects processing to combine the effect signal with the original, clean input signal. This results in a balanced and scaled mixed signal, allowing for a more nuanced control over the effect's intensity and character,
\begin{equation}
y(k)=\mathit{sc}((1-\alpha)x(k)+\alpha y\sub{cnb}(k)),
\label{eq:scaledmix}
\end{equation}
where the index cnb represents the curve number, which can be 1, 2, 3, or 4 in our collection. The parameter $\alpha$ is used to adjust the desired THD value, and the scaling factor $\mathit{sc}$ is used to maintain the power of the mixed signal at the same level as the power of the clean input $x$. For our later feedback experiments, we aimed to use THD values of 5 and 10\,\%. To find the corresponding $\alpha$ values, we used an input sine wave of 400\,Hz with a moderate magnitude of 0.5, as shown in Table\,\ref{tab:thd}. The values for $\mathit{sc}$ will be adjusted later for each individual speech signal. As mentioned earlier, THD is defined for sinusoidal signals only. Our goal is to provide a procedure to assign a "kind of" THD value to a distorted speech signal using the simple and established THD measure, ultimately helping to compare non-linear curves.
\begin{table}
\centering
\caption{Adjusting the desired THD with the help of parameter $\alpha$}
\begin{tabular}{|c|c|c|c|c|}
\hline 
$\alpha$ for curve: & 1 & 2 & 3 & 4 \\ 
\hline 
THD 5\,\% & 0.202 & 0.435 & 0.344 & 0.272 \\ 
\hline 
THD 10\,\% & 0.372 & 0.7 & 0.585 & 0.482 \\ 
\hline 
\end{tabular} 
\label{tab:thd}
\end{table}

In literature, it is often recommended that after applying non-linear processing (e.g., half-wave rectification), the DC level should be removed. However, in our work, we found that including a DC blocker filter may lead to small, but undesired DC offsets in the estimated room impulse responses. Because the DC offset would be small due to the use of small THD values, we decided not to remove the DC from the non-linear processing.

In \cite{4_Valin2016}, it is pointed out that intermodulation caused by non-linearity can have a negative effect. Intermodulation of voiced sounds, like speech (with its harmonics), will produce a harmonic sound, i.e., sum and difference frequencies of integer multiples of the fundamental frequency. We believe that we can neglect the possible negative influence for a low amount of distortion.

Theoretically, prior to the non-linear curve, a high-pass filter should be applied to reduce the signal bandwidth and avoid aliasing from harmonics (from harmonics higher than half the sampling frequency). However, practically, we may expect that the aliasing from the low-level high-frequency harmonics is negligible, and we may omit the additional high-pass filter.

\begin{figure}
\centering
\includegraphics[width=\linewidth]{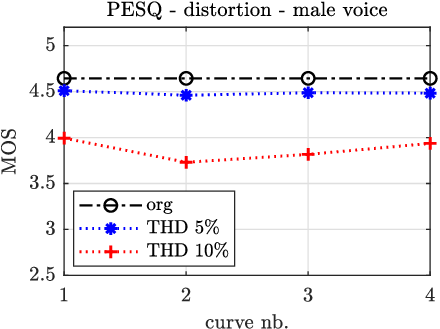}
  \caption{Speech quality MOS for non-linear curve 1 to 4 and $\mathrm{THD}=5$ and 10\,\%.}
  \label{fig:pesq_thd}
\end{figure}

\begin{figure}
\centering
\begin{subfigure}{\linewidth}
  \includegraphics[width=0.9\linewidth]{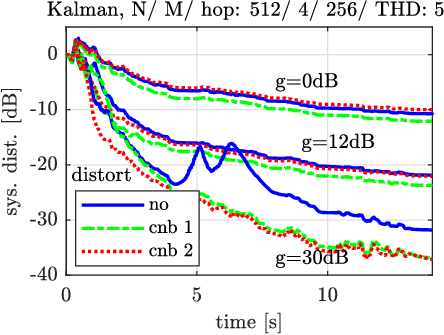}
\end{subfigure}
\begin{subfigure}{\linewidth}
  \includegraphics[width=0.9\linewidth]{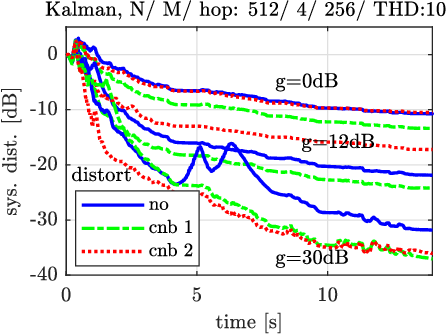}
\end{subfigure} 
\begin{subfigure}{\linewidth}
  \includegraphics[width=0.9\linewidth]{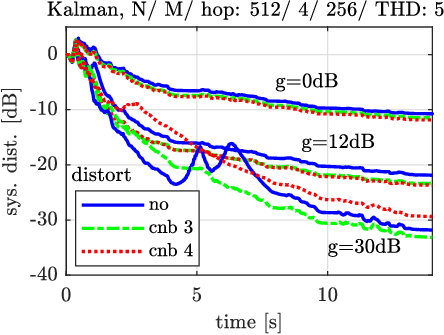}
\end{subfigure} 
\begin{subfigure}{\linewidth}
  \includegraphics[width=0.9\linewidth]{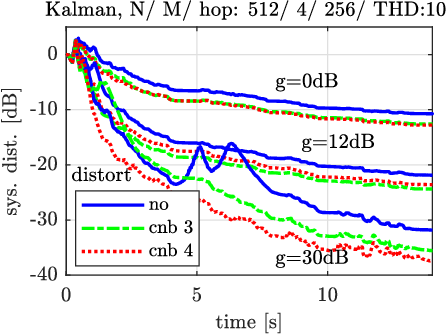}
\end{subfigure} 
  \caption{System distance improvement with non-linear distortion curves 1 to 4 (from top to bottom), $g=0$, 12 and 30\,dB, and $\text{THD}=5$ and 10\,\%.}
  \label{fig:curves_sd}
\end{figure}

For our example male voice, the dependence of MOS on THD values of 5 and 10\,\% is shown in Fig.\,\ref{fig:pesq_thd} (without the feedback system). The $\alpha$ values from Table\,\ref{tab:thd} are used. We found that with THD = 10\,\%, the MOS value is about 4 or already below. THD values between 5 and 10\,\% seem to be a reasonable range.
\noindent
Fig.\,\ref{fig:curves_sd} demonstrates how the addition of distortion curves 1, 2, 3, or 4 stabilizes and improves the system distances. To generate Fig.\,\ref{fig:curves_sd}, we used as a baseline the system with vibrato included, having a maximum delay of $\pm 2$\,msec and a modulation frequency of 1\,Hz, as shown in Fig.\,\ref{fig:vibrato_sd}. Therefore, the distortion is the second extension to the "pure" system. We chose this baseline because the enhancement would be too low if we added distortion only to the "pure" system, and the improvement would not be clearly visible.
\noindent 

\subsection{Bias Reduction with a Reverberation Model}
In \cite{15_Linhard2021}, an energy-decay operator is introduced to build a so-called reverberation model. The basic idea is that the adaptation of the adaptive filter should primarily occur in the reverberant speech parts. For speech onset or double-talk parts, adaptation speed should be reduced. In the reverberant parts, the energy will decay. In parts with energy decay, the adaptation step size should be large. In \cite{15_Linhard2021}, an energy-decay operator (EDO) was introduced (Eq.\,(4) and Eq.\,(8) in \cite{15_Linhard2021}). We will use the form with the ratio of the absolute values of the frequency domain segments of $\vect{X}$
\begin{equation}
\mathrm{EDO}(m,n,l)=\frac{\abs{\vect{X}(n,l-m)}}{\abs{\vect{X}_h(n,l)}}.
\end{equation}
This energy-decay operator will be used to control the adaptation step size, allowing for faster adaptation in reverberant parts and slower adaptation in speech onset or double-talk parts.

Each segment $\vect{X}(n,l)$ is the DFT of half-overlapped time segments (or blocks) of length $N$, with $l$ being the time-discrete block index and $n$ the discrete frequency, and $m$ is the partition index, $m=1,2,\ldots,M$. $\vect{X}_h$ denotes a half-segment, i.e., the first $N/2$ values of the time-segments are zeros. In the application of feedback compensation, we have the special condition that $\vect{X}_h$ is the amplified half-segment error $\vect{E}_h$
\begin{equation}
\vect{X}_h (n,l)= g\cdot \vect{E}_h (n,l).
\end{equation}
This condition allows us to use the energy-decay operator to control the adaptation step size in the feedback compensation system.

For each block time $l$, each frequency $n$, and each partition $m$, we obtain a unique EDO value. However, due to noisy estimates, we limit the EDO range to prevent excessive adaptation step sizes. This is achieved by applying the restriction
\begin{equation}
\mathrm{EDO}=\max\{r\sub{min},\min\{r\sub{max},\mathrm{EDO}\}\},
\label{eq:restrict_edo}
\end{equation}
with, for example, $r_{\min} = 0.2$ and $r_{\max} = 2$. In \cite{15_Linhard2021}, this EDO was used to multiply the standard step-size of an adaptive multi-delay frequency domain least-mean squares filter, MDF-FLMS. Ultimately, a sign-sign adaptation filter (adaptation with phase only) could be derived. 

As an extension to \cite{15_Linhard2021}, we propose a new noise-reduced EDO form, referred to as the "best fit of 2 vectors" or "best curve fit". This approach involves finding the least squares solution to overdetermined systems, which can be achieved using linear algebra operations. However, we will omit the underlying linear algebra details.

In MATLAB notation, for two column vectors, the notation is '\textbackslash' (for row vectors, use '/'). We obtain
\begin{equation}
\mathrm{EDO}(m,n,l)=\abs{\vect{X}(n,l-m)}\textbackslash\abs{\vect{X}_h(n,l)}.
\label{eq:edo_elegant}
\end{equation}
The result is a scalar, representing the best fit, which is a function depending on frequency $n$. Alternatively, a mathematically simpler version can be used, involving the ratio of the mean values
\begin{equation}
\mathrm{EDO}(m,l)=\frac{\mathrm{mean}\{\abs{\vect{X}(n,l-m)}\}}{\mathrm{mean}\{\abs{\vect{X}_h(n,l)}\}}.
\label{eq:edo_simple}
\end{equation}
This simpler version calculates the mean over frequency. We can restrict the EDO range according to Eq.\,\ref{eq:restrict_edo} and multiply the standard Kalman step-size $\alpha$ with this EDO factor.

\begin{figure}
\centering
\begin{subfigure}{\linewidth}
  \includegraphics[width=\linewidth]{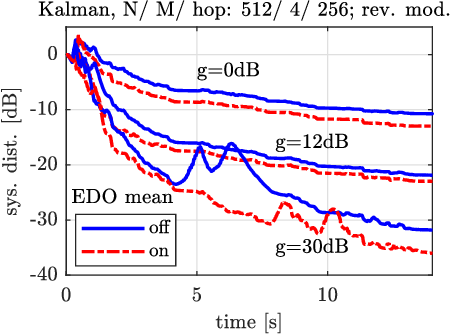}
\end{subfigure}
\begin{subfigure}{\linewidth}
  \includegraphics[width=\linewidth]{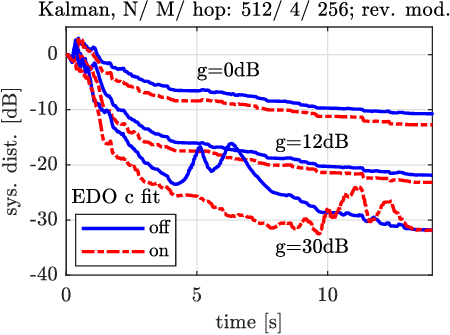}
\end{subfigure} 
  \caption{Including a reverberation model based on the ratio of mean values (top) and curve fit (bottom) with gains 0, 12 or 30\,dB.}
  \label{fig:edo}
\end{figure}

For our example of male voice, we conducted a first comparison of the mathematically elegant EDO version Eq.\,\ref{eq:edo_elegant} and the very simple EDO version Eq.\,\ref{eq:edo_simple}, as shown in Fig.\,\ref{fig:edo}. Both versions exhibit a similar improved performance, measured as a smaller system distance. Similar to Fig.\,\ref{fig:curves_sd}, we needed to use the EDO as a 2nd addition to get a clearly visible improvement, i.e., the baseline system includes vibration with a maximum delay of $\pm 2$\,msec and 1\,Hz modulation. We can directly compare the results to Fig.\,\ref{fig:curves_sd}, which shows the results for $\mathrm{THD} = 5$\,\% and $\mathrm{THD} = 10$\,\%.

\noindent 

\section{Increasing the Convergence Speed}
\label{sec:convergence_speed}
\begin{figure}
  \centering
  \includegraphics{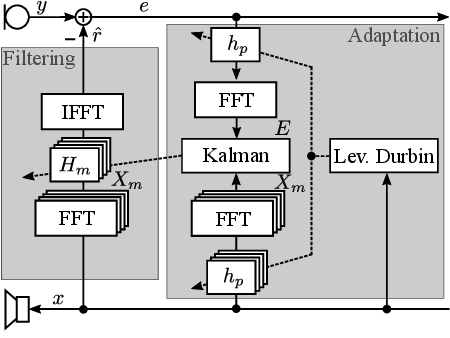}
  \caption{Filtering and adaptation part of the feedback system with prediction applied for adaptation only.}
  \label{fig:lev_durbin}
\end{figure}

The first step in increasing the convergence speed is the whitening that results from the frequency domain approach, which we have already mentioned in the introduction. A further important step is placing a prediction filter block at the input signal $x$. However, prediction filtering should only influence the filter adaptation and not the true signal path. Fig.\,\ref{fig:lev_durbin} illustrates that filter adaptation can be seen as a separate block parallel to the main filter. The filter coefficients are copied from the adaptation block to the filtering block. We can find these extended structures in literature, such as in \cite{10_Kuehl2017,11_Bernardi2017}.
\noindent

The estimation of the predictors $\h_p$, a vector of length $N_p$, can be performed using the standard method of linear predictive coding (LPC) with the Levinson-Durbin algorithm. $N_p$ is the predictor order. Our type of parallel structure is also shown in, for example, \cite{10_Kuehl2017}. However, in contrast to \cite{10_Kuehl2017}, we use a multi-delay structure where the input $x$ is divided into several overlapping parts. In the frequency domain, $\vect{X}(m,l)$ is a vector of length $N$. In the MDF structure, $M$ is the number of partitions, $m=1, 2,\ldots, M$, and $l$ is the block time index. We will choose $M$ to achieve a block rate of about 20\,msec, which is the usual value for speech LPC calculations. For each of the overlapped segments of $\vect{X}(m,l)$, we calculate an own predictor $\h_p(l-m+1)$. For example, at the new block index $l$ and $m=1$, the new input $\vect{X}(1,l)$ requires a new calculation of $\h_p(l)$, but for $m=2$, we already have $\vect{X}(2,l)$ with the already calculated predictor $\h_p(l-1)$, and so on for the other $m$ values. Synchronizing the calculation of the predictor to the segment rate leads to the inherent problem that $M$ different predictors contribute to one adaptation step. If we were to use $M$ identical predictors for one adaptation step, this predictor would not be the best match for all $M$ different partitions $\vect{X}(m,l)$. We suggest two compromise solutions:
\begin{itemize}
\item[A)] Use the newest predictor $\h_p(l)$ for all $M$ partitions.
\item[B)] Calculate a common predictor for always two partitions, resulting in $M/2$ different predictors involved with one adaptation step.
\end{itemize}
The extra calculation amount for prediction B) is about 1/2 compared to A).

\begin{figure}[t]
\centering
\begin{subfigure}{\linewidth}
  \includegraphics[width=\linewidth]{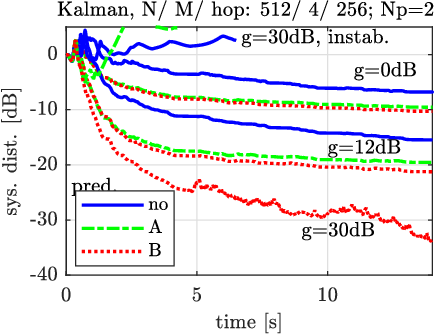}
\end{subfigure}
\begin{subfigure}{\linewidth}
  \includegraphics[width=\linewidth]{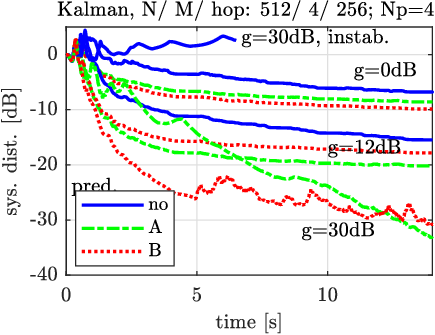}
\end{subfigure} 
  \caption{Including prediction into the adaptation part with schemes A) and B) for $g=0$, 12, 30\,dB, top: predictor order 2; bottom: order 4.}
  \label{fig:lpc}
\end{figure}

Fig.\,\ref{fig:lpc} depicts the convergence of the system distance for loop gains 0, 12, and 30\,dB for the prediction update schemes A) and B) compared to the case with no prediction. The improvement with prediction shown in Fig.\,\ref{fig:lpc} can be directly compared to the improvement obtained with the vibrato, as shown in Fig.\,\ref{fig:vibrato_sd}. The LPC order $N_p=2$ or 4 is used. Using higher orders did not significantly improve the results (a low order was already suggested in \cite{10_Kuehl2017}).

\begin{figure}[t]
\centering
\begin{subfigure}{\linewidth}
  \includegraphics[width=\linewidth]{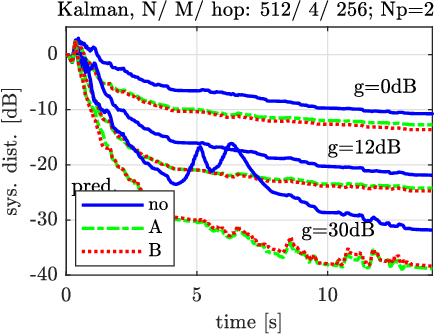}
\end{subfigure}
\begin{subfigure}{\linewidth}
  \includegraphics[width=\linewidth]{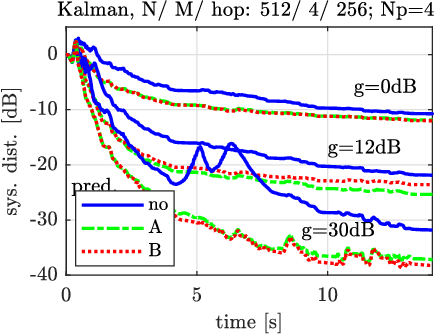}
\end{subfigure} 
  \caption{Including vibrato (base line) and prediction with schemes A) and B) for $g=0$, 12, 30\,dB, top: predictor order 2; bottom: order 4.}
  \label{fig:lpc_vibrato}
\end{figure}

\noindent 
For a comparison to distortion and EDO, as shown in Fig.\,\ref{fig:curves_sd} and Fig.\,\ref{fig:edo}, we present Fig.\,\ref{fig:lpc_vibrato}. In this figure, prediction is already the 2nd addition, with the vibrato (max. delay $\pm 2$\,msec, 1\,Hz modulation, baseline) as the 1st addition.
\noindent 
There is almost no difference between schemes A) and B). 

\section{Combination of Several Decorrelation Methods and Data-base Results}
\label{sec:combination_decorrelation}
We combine several decorrelation methods and evaluate the complete system using data from two public databases. The speech data is taken from a Lombard speech database in German language \cite{18_Soloducha2016}, and the impulse responses are obtained from the ANIR (Automotive Noise and Impulse Response) corpus \cite{19_Huebschen2022}. This dataset is identical to the one used in \cite{16_Linhard2025}. From the speech database, we selected two female and two male speakers, each with two sentences. We only used the Lombard-free speech, as our focus is not on the Lombard effect. Since each speech sentence has a length of about 6 to 10 seconds, we repeated each sentence to create a longer sequence of about 42 seconds. From the ANIR corpus, we used three different impulse responses, specifically the impulse response from the headliner driver microphone (no. 1) to either the door speakers of the driver or the left or right side door loudspeakers in the car fond (nos. 18, 20, and 21). We performed a simple low-frequency equalization, as the original impulse responses showed too much bass. We resampled all the data to 16 kHz. The feedback core system is the adaptive Kalman filter with the structure and parameters previously described. The acoustic coupling between the signal from the loudspeaker $r$ and the original speech signal $s$ is roughly about -10 dB.

The combinations of eight sentences with three impulse responses yield 24 speech samples. We use four gain settings for the feedback loop, $g=0$, 6, 12, and 30 dB. We found that results from the three impulse responses could be averaged together, but we should average male and female voices separately. In this way, for each of the four gain values, we get an averaged result of 12 male or 12 female speech sentences.

The quality metrics used to evaluate the system are a combination of:
\begin{itemize}
\item PESQ, MOS value
\item Early system distance, $\mathit{sd}_5$
\item Late system distance, $\mathit{sd}_{20+}$
\item Overflow rate (hard limiter with 6\,dB head room) 
\end{itemize}
The concept of early and late system distances to indicate convergence speed was introduced in Sec.\,\ref{sec:performance} and \cite{16_Linhard2025}. The MOS value is calculated from the last complete sentence, specifically the last complete sentence in the interval $[20\ 41]$\,sec. For this second part of the evaluation, which uses databases and automatic evaluation, we included a hard limiter, as discussed in Sec.\,\ref{sec:performance}. If overflow occurred, it mainly happened for the case $g=30$\,dB. For the following presentation, we will only note the overflow rate for this case, and we will express it as a percentage [\%].

\begin{figure*}
    \centering
    
    \caption*{Baseline, no decorrelation}
    \begin{subfigure}{0.45\linewidth}
        \includegraphics[width=\linewidth]{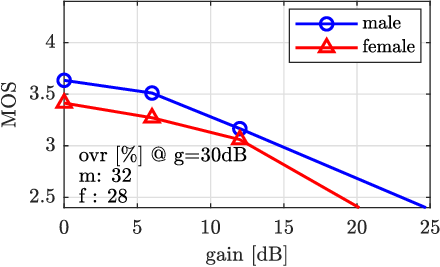}
    \end{subfigure}
    \begin{subfigure}{0.45\linewidth}
        \includegraphics[width=\linewidth]{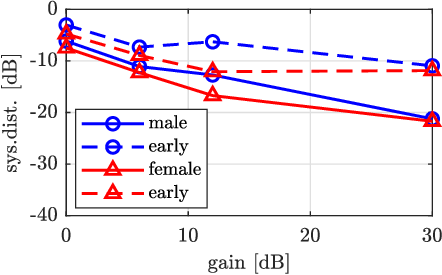}
    \end{subfigure}
    
    \caption*{Reverb mod. (curve fitting) only}
    \begin{subfigure}{0.45\linewidth}
        \includegraphics{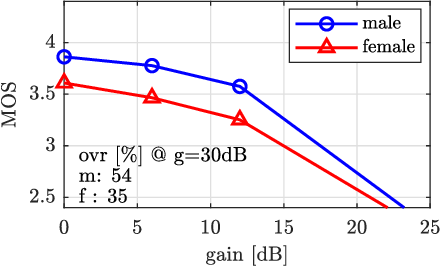}
    \end{subfigure}
    \begin{subfigure}{0.45\linewidth}
        \includegraphics{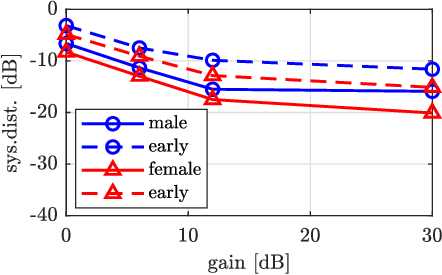}
    \end{subfigure}
    
    \caption*{Distortion ($\text{THD}=5$\,\%) only}
    \begin{subfigure}{0.45\linewidth}
        \includegraphics{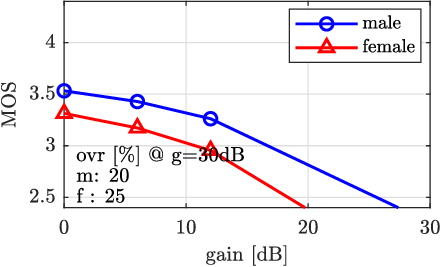}
    \end{subfigure}
    \begin{subfigure}{0.45\linewidth}
        \includegraphics{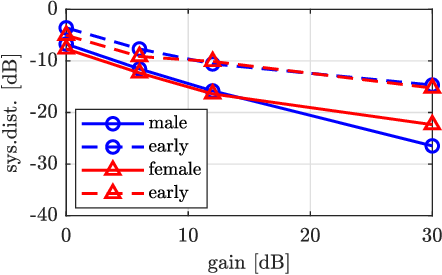}
    \end{subfigure}
    
    \caption*{Reverb mod. and distortion}
    \begin{subfigure}{0.45\linewidth}
        \includegraphics{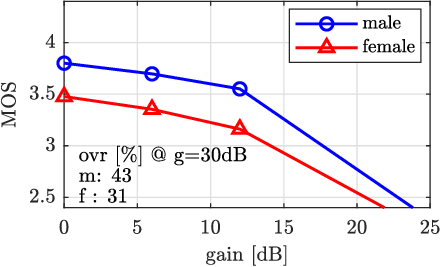}
    \end{subfigure}
    \begin{subfigure}{0.45\linewidth}
        \includegraphics{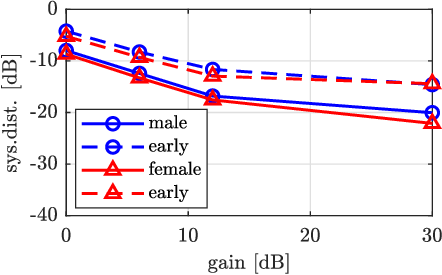}
    \end{subfigure}
    
    \caption{Comparison of 4 feedback systems with 4 extensions, from top to bottom: baseline, reverb model only, distortion only, reverb model \& distortion combined.}
	\label{fig:comparison1}
\end{figure*}\noindent

In Fig.\,\ref{fig:comparison1}, the first row presents the baseline result. For a gain of $g=0$\,dB, PESQ yields 3.6 or 3.4 for male and female speech, respectively. However, for $g=30$\,dB, PESQ exhibits poor MOS values, which are below 2.4 and thus outside the range of the figure. The overflow rate for $g=30$\,dB is 32 or 28\,\% for male and female speech, respectively. The system distance for the highest gain reaches about -10 or -20\,dB for male and female speech, respectively. We can conclude that the system performance would be too poor for a practical system.

The 2nd and 3rd rows of Fig.\,\ref{fig:comparison1} display some small improvements obtained with the simple decorrelation techniques, namely the reverb model (case: curve fitting) and distortion (case: $\mathrm{THD}=5$\,\%). The 4th row presents the results for the combination of reverb model and distortion.

We observed that the reverb model improves PESQ, whereas distortion reduces PESQ but enhances the system distance. A first optimum may be the combination of these two simple decorrelation techniques, although the improvements are not very large.

\begin{figure*}
    \centering
    
    \caption*{Prediction only}
    \begin{subfigure}{0.45\linewidth}
        \includegraphics[width=\linewidth]{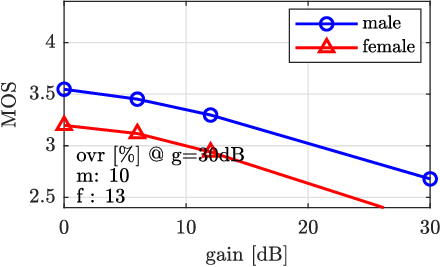}
    \end{subfigure}
    \begin{subfigure}{0.45\linewidth}
        \includegraphics[width=\linewidth]{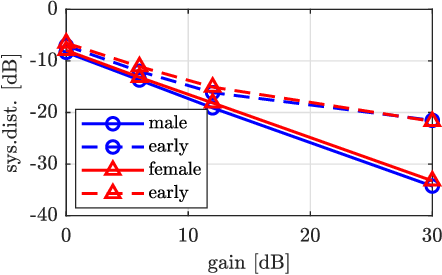}
    \end{subfigure}

    \caption*{Vibrato}
    \begin{subfigure}{0.45\linewidth}
        \includegraphics{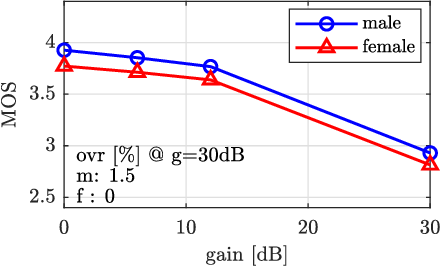}
    \end{subfigure}
    \begin{subfigure}{0.45\linewidth}
        \includegraphics{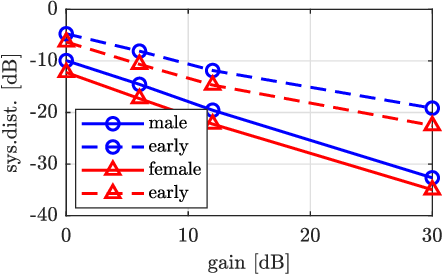}
    \end{subfigure}

    \caption*{Prediction and vibrato}
    \begin{subfigure}{0.45\linewidth}
        \includegraphics{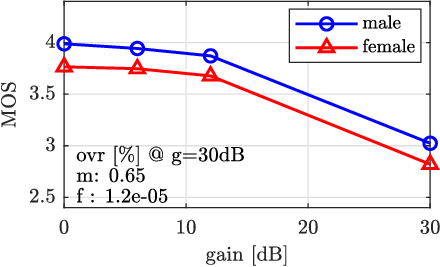}
    \end{subfigure}
    \begin{subfigure}{0.45\linewidth}
        \includegraphics{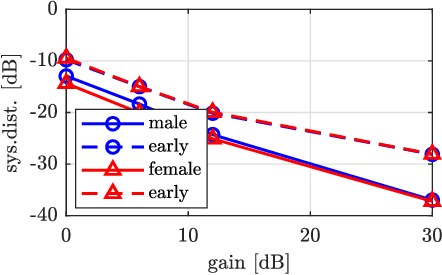}
    \end{subfigure}
    
    \caption*{Pred. and vib. and dist. and rev}
    \begin{subfigure}{0.45\linewidth}
        \includegraphics{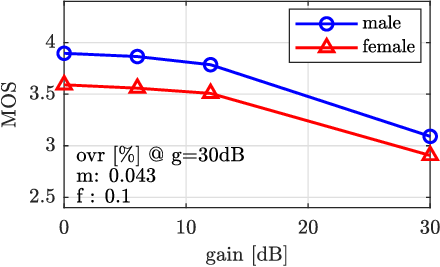}
    \end{subfigure}
    \begin{subfigure}{0.45\linewidth}
        \includegraphics{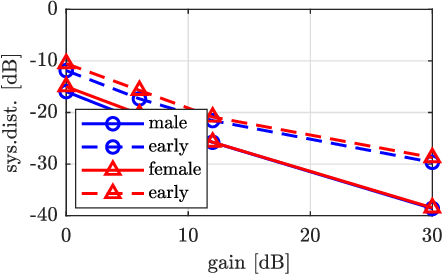}
    \end{subfigure}
    
    \caption{Comparison of 4 feedback systems with 4 extensions, from top to bottom: prediction only, vibrato only, pred. + vib ., pred. + vib. + distortion + reverb model.}
	\label{fig:comparison2}
\end{figure*}\noindent

In Fig.\,\ref{fig:comparison2}, the top comparison shows a system extended with prediction only. In this case, the overflow rate at 30\,dB (ovr@30dB) is 10\,\% for male voices and 13\,\% for female voices. However, PESQ significantly increases in the next row of Fig.\,\ref{fig:comparison2}, which corresponds to the case vibrato only. The variable delay line (vibrato) parameters are: frequency = 1\,Hz, maximum delay = 2\,msec. The overflow rate for male voices is only 1.5\,\% and 0\,\% for female voices.

The 3rd system (3rd row shown in Fig.\,\ref{fig:comparison2}) combines the predictor and variable delay line. Again, performance is improved. Finally, for the last system in this comparison, we additionally extended the 3rd system with the reverberation model and a 5\,\% distortion. Thus, we have integrated all decorrelation methods discussed in the previous sections. A small advantage for the system distances appears, but PESQ gets slightly worse due to the distortion. The overflow rate at 30\,dB is in the range of 0.1\,\% or less.

A remarkable result is that the prediction extension does not improve the PESQ result, although the system distance is improved. However, the combination of prediction and vibrato seems to provide an ideal combination, with both PESQ and system distance improved.

\section{Conclusion}
We discussed an acoustic feedback system based on a frequency domain Kalman filter in a multi-delay structure to model a long acoustic path of 1000 or more samples. The loop gain is in the range of 0 to 30\,dB. The basic system without any extensions shows very poor performance. We first examined the isolated extensions and quantified the improvements, respectively. In several steps, we included the extensions, one after the other, and finally all methods together. The final evaluation on a speech and impulse response database confirmed the improvements.

Introducing distortion into the system always resulted in a measurable decrease of speech quality, i.e., a reduced MOS value. We fixed the amount of distortion to THD 5\,\% to guarantee only a small MOS reduction. For the reverberation model, we decided finally for the very simple curve fitting (solution of an overdetermined system). It helps stabilize the estimation process. For prediction, a low prediction order of e.g., 2 seems to be enough. For vibrato, a modulation delay of max. 2\,msec (about 1 to 2\,Hz modulation frequency) gives a sufficient improvement without affecting speech quality.

Comparing the single methods, the time-variable delay line (vibrato) gave the most significant improvement, followed by prediction. Combining the two effects, vibrato and prediction, already gives a superior improved system, better than any single extension. The further inclusion of distortion and a simple reverberation model may further improve the total performance but only to a small extent.

\section*{Data Availability Statement}
The audio and impulse response data used in this work come from publicly available resources. The Lombard speech recordings \cite{18_Soloducha2016} are available on Zenodo (\url{https://zenodo.org/records/48713}). The ANIR in-car impulse response corpus \cite{19_Huebschen2022} is available from the Digital Signal Processing and System Theory Group at Kiel University (\url{https://dss-kiel.de/index.php/media-center/data-bases/anir-corpus}). All datasets, used for the final evaluation, are accessible to the public under the terms specified by their respective providers. No proprietary or restricted data were used. 

\bibliography{iccBib} 

@INPROCEEDINGS{1_Benesty1997,
  author={Benesty, J. and Morgan, D.R. and Sondhi, M.M.},
  booktitle={1997 IEEE International Conference on Acoustics, Speech, and Signal Processing}, 
  title={A better understanding and an improved solution to the problems of stereophonic acoustic echo cancellation}, 
  year={1997},
  volume={1},
  number={},
  pages={303-306 vol.1},
  doi={10.1109/ICASSP.1997.599629}}

@ARTICLE{2_Morgan2001,
  author={Morgan, D.R. and Hall, J.L. and Benesty, J.},
  journal={IEEE Transactions on Speech and Audio Processing}, 
  title={Investigation of several types of nonlinearities for use in stereo acoustic echo cancellation}, 
  year={2001},
  volume={9},
  number={6},
  pages={686-696},
  doi={10.1109/89.943346}}

@misc{4_Valin2016,
      title={Channel Decorrelation For Stereo Acoustic Echo Cancellation In High-Quality Audio Communication}, 
      author={Jean-Marc Valin},
      year={2016},
      note={arXiv:1603.03364 [cs.SD], \url{https://arxiv.org/abs/1603.03364}}, 
}

@Article{5_Soo1990,
  Title                    = {Multidelay Block Frequency Domain Adaptive Filter},
  Author                   = {Soo, Jia-Sien and Pang, Khee K.},
  Journal                  = {IEEE Transactions on Acoustics, Speech, and Signal Processing},
  Year                     = {1990},

  Month                    = feb,
  Number                   = {2},
  Pages                    = {373-376},
  Volume                   = {38}
}

@article{6_Enzner2006,
title = {Frequency-domain adaptive Kalman filter for acoustic echo control in hands-free telephones},
journal = {Signal Processing},
volume = {86},
number = {6},
pages = {1140-1156},
year = {2006},
note = {Applied Speech and Audio Processing},
issn = {0165-1684},
doi = {https://doi.org/10.1016/j.sigpro.2005.09.013},
url = {https://www.sciencedirect.com/science/article/pii/S0165168405003178},
author = {Gerald Enzner and Peter Vary},
}

@INPROCEEDINGS{7_Kuech2014,
  author={K{\"u}ch, Fabian and Mabande, Edwin and Enzner, Gerald},
  booktitle={2014 IEEE International Conference on Acoustics, Speech and Signal Processing (ICASSP)}, 
  title={State-space architecture of the partitioned-block-based acoustic echo controller}, 
  year={2014},
  volume={},
  number={},
  pages={1295-1299},
  doi={10.1109/ICASSP.2014.6853806}}

@INPROCEEDINGS{8_Puder2017,
  author={Puder, Henning and Strasser, Falco},
  booktitle={2017 25th European Signal Processing Conference (EUSIPCO)}, 
  title={Decorrelation measures for stabilizing adaptive feedback cancellation in hearing aids}, 
  year={2017},
  volume={},
  number={},
  pages={583-587},
  doi={10.23919/EUSIPCO.2017.8081274}}

@INPROCEEDINGS{9_Guo2012,
  author={Guo, Meng and Jensen, Søren Holdt and Jensen, Jesper and Grant, Steven L.},
  booktitle={2012 Proceedings of the 20th European Signal Processing Conference (EUSIPCO)}, 
  title={On the use of a phase modulation method for decorrelation in acoustic feedback cancellation}, 
  year={2012},
  volume={},
  number={},
  pages={2000-2004},
  doi={}}

@INPROCEEDINGS{10_Kuehl2017,
  author={K{\"u}hl, Stefan and Antweiler, Christiane and Hübschen, Tobias and Jax, Peter},
  booktitle={2017 Hands-free Speech Communications and Microphone Arrays (HSCMA)}, 
  title={Kalman filter based stereo system identification with auto- and cross-decorrelation}, 
  year={2017},
  volume={},
  number={},
  pages={181-185},
  doi={10.1109/HSCMA.2017.7895586}}

@ARTICLE{11_Bernardi2017,
  author={Bernardi, Giuliano and van Waterschoot, Toon and Wouters, Jan and Moonen, Marc},
  journal={IEEE/ACM Transactions on Audio, Speech, and Language Processing}, 
  title={Adaptive Feedback Cancellation Using a Partitioned-Block Frequency-Domain Kalman Filter Approach With PEM-Based Signal Prewhitening}, 
  year={2017},
  volume={25},
  number={9},
  pages={1784-1798},
  doi={10.1109/TASLP.2017.2716188}}

@Book{12_Zoelzer2011,
  author    = {Z{\"o}lzer, Udo},
  publisher = {John Wiley and Sons Ltd},
  title     = {DAFX: Digital Audio Effects},
  year      = {2011},
  edition   = {2},
}

@InProceedings{13_Smith2022,
  Title					  = {Interpolated Delay Lines, Ideal Bandlimited Interpolation, and Fractional Delay Filter Design},
  Author			      = {Smith III, Julius Orion},
  Booktitle				  = {MUS420 Lecture 4a},
  Year					  = {2022}
}

@InCollection{14_Schmidt2006,
  Title                    = {Signal Processing for In-Car Communication Systems},
  Author                   = {Schmidt, Gerhard and Haulick, Tim},
  Booktitle                = {Topics in Acoustic Echo and Noise Control},
  Publisher                = {Springer},
  Year                     = {2006},

  Address                  = {Berlin},
  Chapter                  = {14},
  Editor                   = {H{\"a}nsler, E. and Schmidt, G.},
  Pages                    = {437-493}
}

@InProceedings{15_Linhard2021,
  Title					  = {Robust and High Gain Acoustic Feedback Compensation in the Frequency Domain With a Simple Energy-decay Operator},
  Author					  = {Linhard, Klaus and Bulling, Philipp and  Gimm, Marco and Schmidt, Gerhard},
  Booktitle				  = {14th ITG Conference on Speech Communication},
  Year					  = {2021}
}

@misc{16_Linhard2025,
      title={A Phase Synthesizer for Decorrelation to Improve Acoustic Feedback Cancellation}, 
      author={Klaus Linhard and Philipp Bulling},
      year={2025},
      note={arXiv:2510.12377 [eess.AS] \url{https://arxiv.org/abs/2510.12377}}, 
}

@ARTICLE{17_Hu2008,
  author={Hu, Yi and Loizou, Philipos C.},
  journal={IEEE Transactions on Audio, Speech, and Language Processing}, 
  title={Evaluation of Objective Quality Measures for Speech Enhancement}, 
  year={2008},
  volume={16},
  number={1},
  pages={229-238},
  doi={10.1109/TASL.2007.911054}}

@InProceedings{18_Soloducha2016,
  Title                    = {Lombard speech database for German Language},
  Author                   = {Soloduca, Michal and Raake, Alexander and Kettler, Frank and Voigt, Peter},
  Booktitle                = {42. Deutsche Jahrestagung f{\"u}r Akustik (DAGA)},
  Year                     = {2016},

  Address                  = {Aachen, Deutschland},
  Month                    = mar
}

@InProceedings{19_Huebschen2022,
  Title                    = {A Background Noise and Impulse Response Corpus for Research in Automotive Speech and Audio Processing},
  Author                   = {H{\"u}bschen, Tobias and Gimm, Marco and Schmidt, Gerhard},
  Booktitle                = {48. Deutsche Jahrestagung f{\"u}r Akustik (DAGA)},
  Year                     = {2022},

  Address                  = {Stuttgart, Deutschland},
  Month                    = mar
}

@InProceedings{Bulling2016,
  Title                    = {{Nichtlineare Kennlinien zur Verbesserung der Sprachverst{\"a}ndlichkeit in ger{\"a}uschbehafteter Umgebung}},
  Author                   = {Bulling, Philipp and Linhard, Klaus and Wolf, Arthur and Schmidt, Gerhard and Thei{\ss{}}, Anne and Gimm, Marco},
  Booktitle                = {42. Deutsche Jahrestagung f{\"u}r Akustik (DAGA)},
  Year                     = {2016},

  Address                  = {Aachen, Deutschland},
  Month                    = mar
}
\bibliographystyle{ieeetr}

\end{document}